\documentclass[letterpaper,twocolumn]{revtex4-1}
\usepackage{graphicx}
\usepackage{color}
\usepackage{amsmath,amssymb,amsfonts} 

\def\d#1{{\rm d}\kern -0.2mm#1}
\def\fig{Fig.}
\def\figs{Figs.}
\def\eq{Eq.}

\usepackage{srcltx}
\begin{document}

\title{Real-Time Scanning Charged-Particle Microscope Image Composition with Correction of Drift}
\author{Petr Cizmar, Andr\'as E. Vlad\' ar, and Michael T. Postek}
\affiliation{National Institute of Standards and Technology\footnote{
Contribution of the National Institute of Standards and Technology; not subject
to copyright. Certain commercial equipment is identified in this report to
adequately describe the experimental procedure. Such identification does not
imply recommendation or endorsement by the National Institute of Standards and
Technology, nor does it imply that the equipment identified is necessarily the
best available for the purpose.}, 100 Bureau Drive, Gaithersburg, MD 20899}

\begin{abstract}
In this article, a new scanning electron microscopy (SEM) image composition technique is
described, which can significantly reduce drift related image corruptions. 
Drift-distortion commonly causes blur and distortions in the SEM
images. Such corruption ordinarily appears when conventional
image-acquisition methods, i.e. ``slow scan'' and ``fast scan'', are applied.
The damage is often very significant; it may render images unusable for
metrology applications, especially, where sub-nanometer accuracy is required. The
described correction technique works with a large number of quickly taken frames,
which are properly aligned and then composed into a single image. Such image
contains much less noise than the individual frames, whilst the blur and
deformation is minimized. This technique also provides useful information about
changes of the sample position in time, which may be applied to investigate
the drift properties of the instrument without a need of additional equipment.
\end{abstract}
\maketitle

\begin{figure}
(a)\includegraphics[width=6cm]{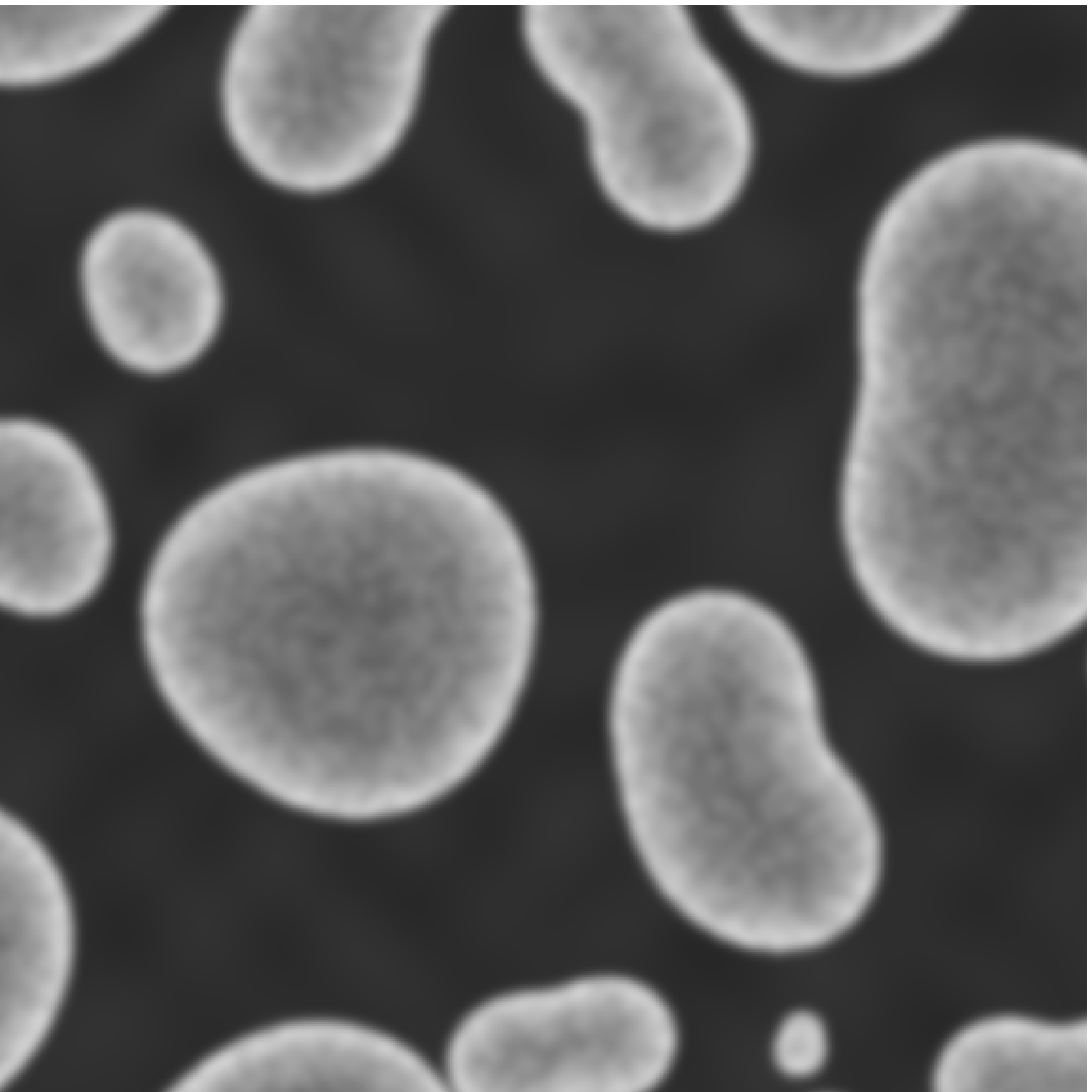}
(b)\includegraphics[width=6cm]{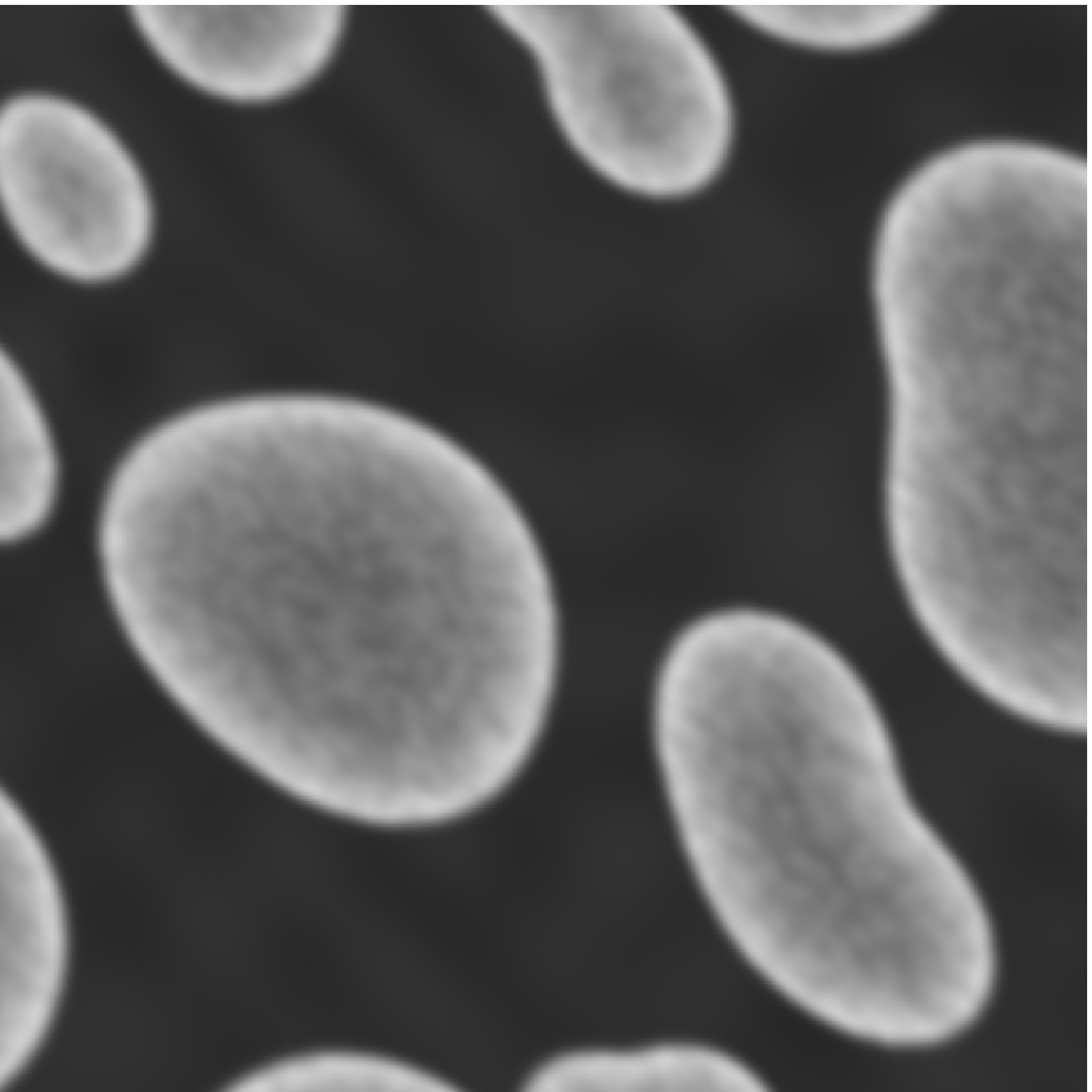}\\
\caption{
Illustration of drift-distortion-related image corruption on simulated
\cite{cizmar-simim-scanning} ``slow
scan'' SEM images of a gold-on-carbon
sample. (a)~Ideal, undistorted image. (b)~Typical corrupted image.}
\label{fast_composition_fig_examples}
\end{figure}

\section*{Introduction}
Advances in fundamental nano-science, development of nano-materials, and
eventually manufacturing of nanometer-scale products all depend to some extent
on the capability to accurately and reproducibly measure dimensions, properties,
and performance characteristics at the nano-scale. Scanning electron microscopes
(SEMs) have been used in this application for many years \cite{postek-advanced}, \cite{postek-photomask}.  Since progress in
nano-science and nano-technology has been rather rapid recently, the
dimensions of nano-structures and nano-objects have shrunk significantly.
Consequently, accurate SEM imaging has been emphasized. The dimensions or
distances have been measured from SEM images or line-scans.
Current imaging methods in SEM are often incapable of achieving the desired accuracy,
because the SEM images, at such high magnifications, often suffer from
drift-related
distortion. In  many cases, the drift is significant and the SEM images exhibit
deformations or blur. The same problem is also experienced in other fields, e.g.
scanning probe microscopies.

\begin{figure}
(a)\includegraphics[width=6cm]{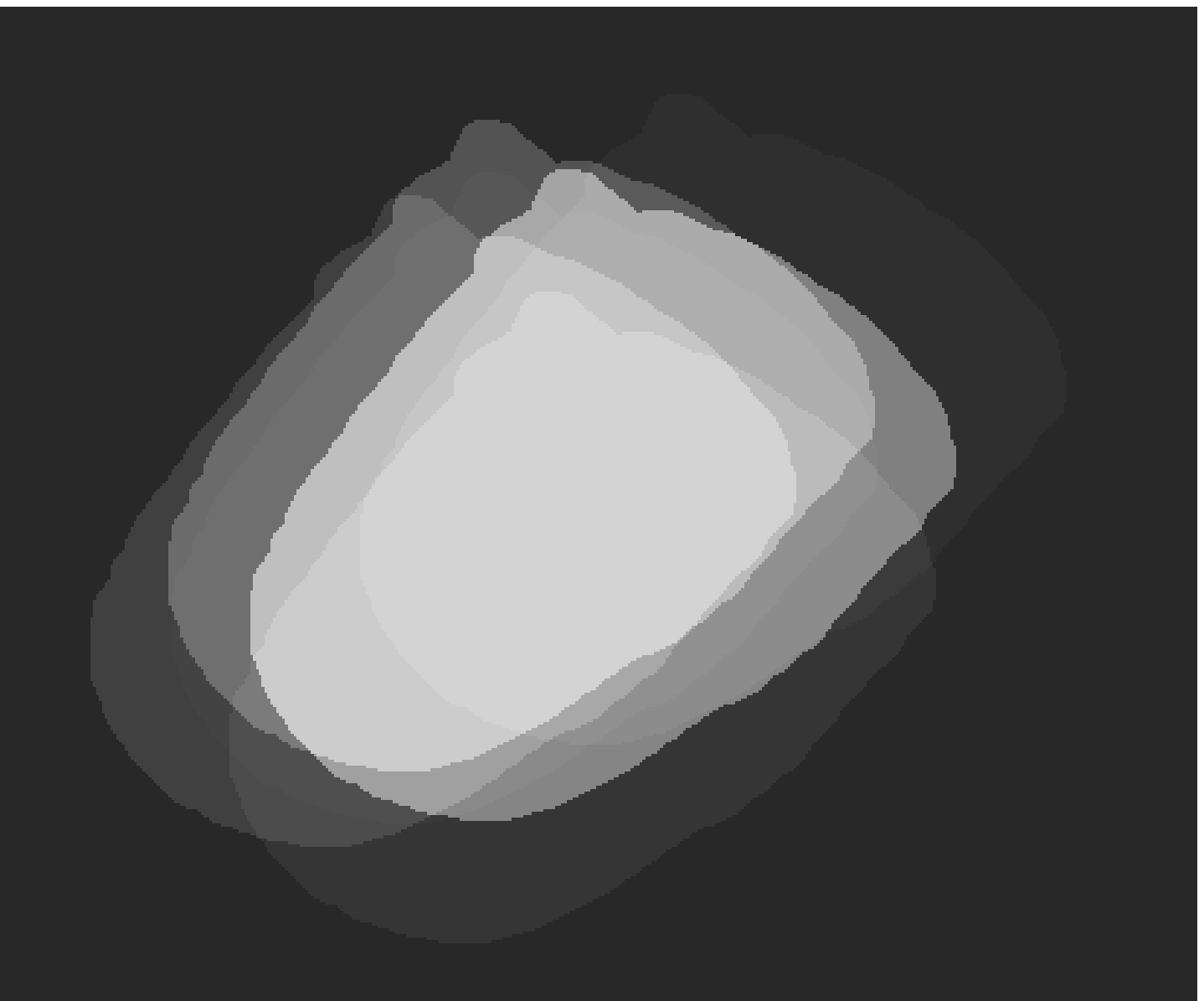}
(b)\includegraphics[width=6cm]{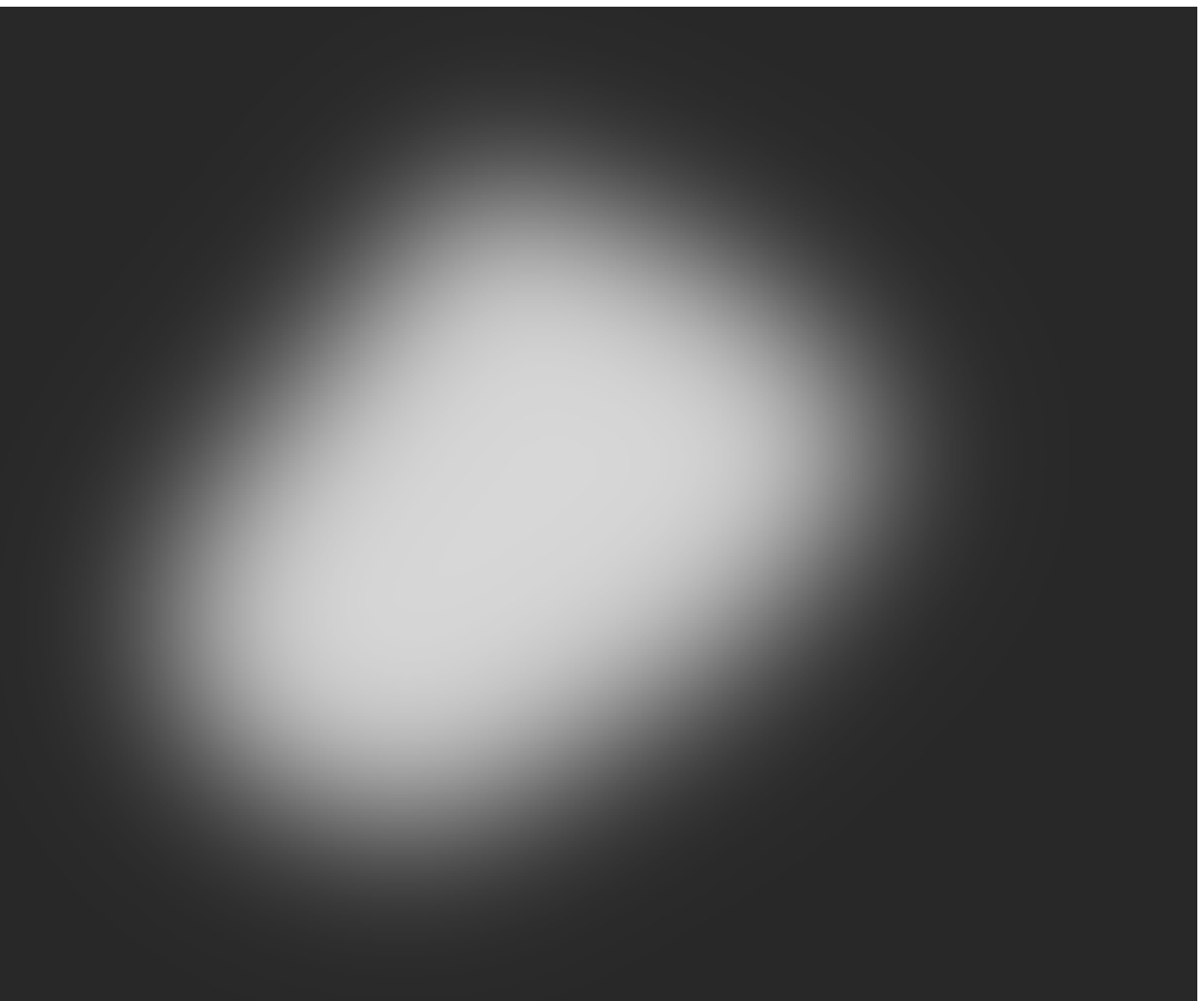}\\
\caption{
Illustration of the composition (averaging) of displaced image frames.
(a)~Composition of a few image frames,
(b)~composition of a large number of frames, the image exhibits excessive blur.
}
\label{fast_composition_fig_shift_demo}
\end{figure}

\begin{figure}
(a)\includegraphics[width=6cm]{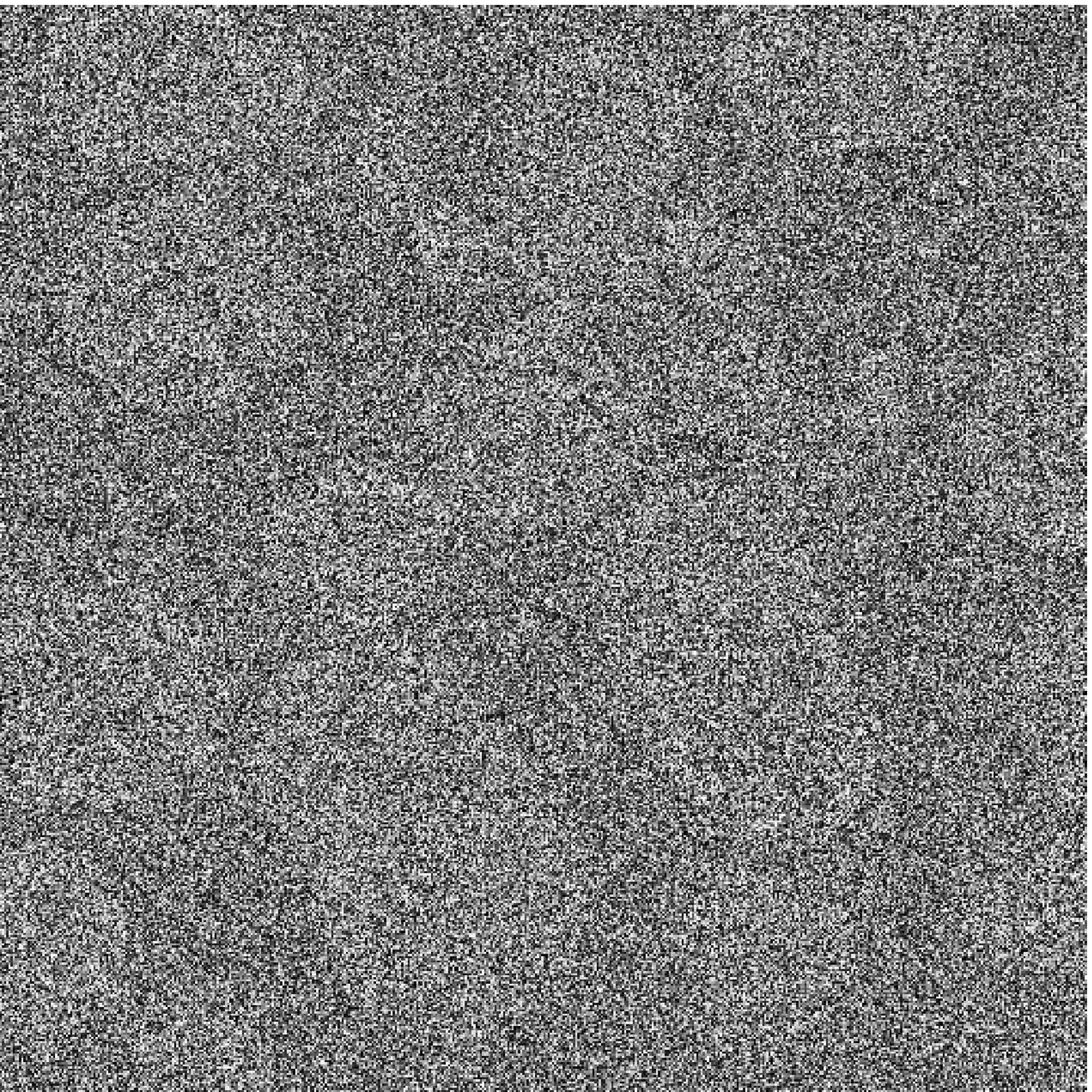}
(b)\includegraphics[width=6cm]{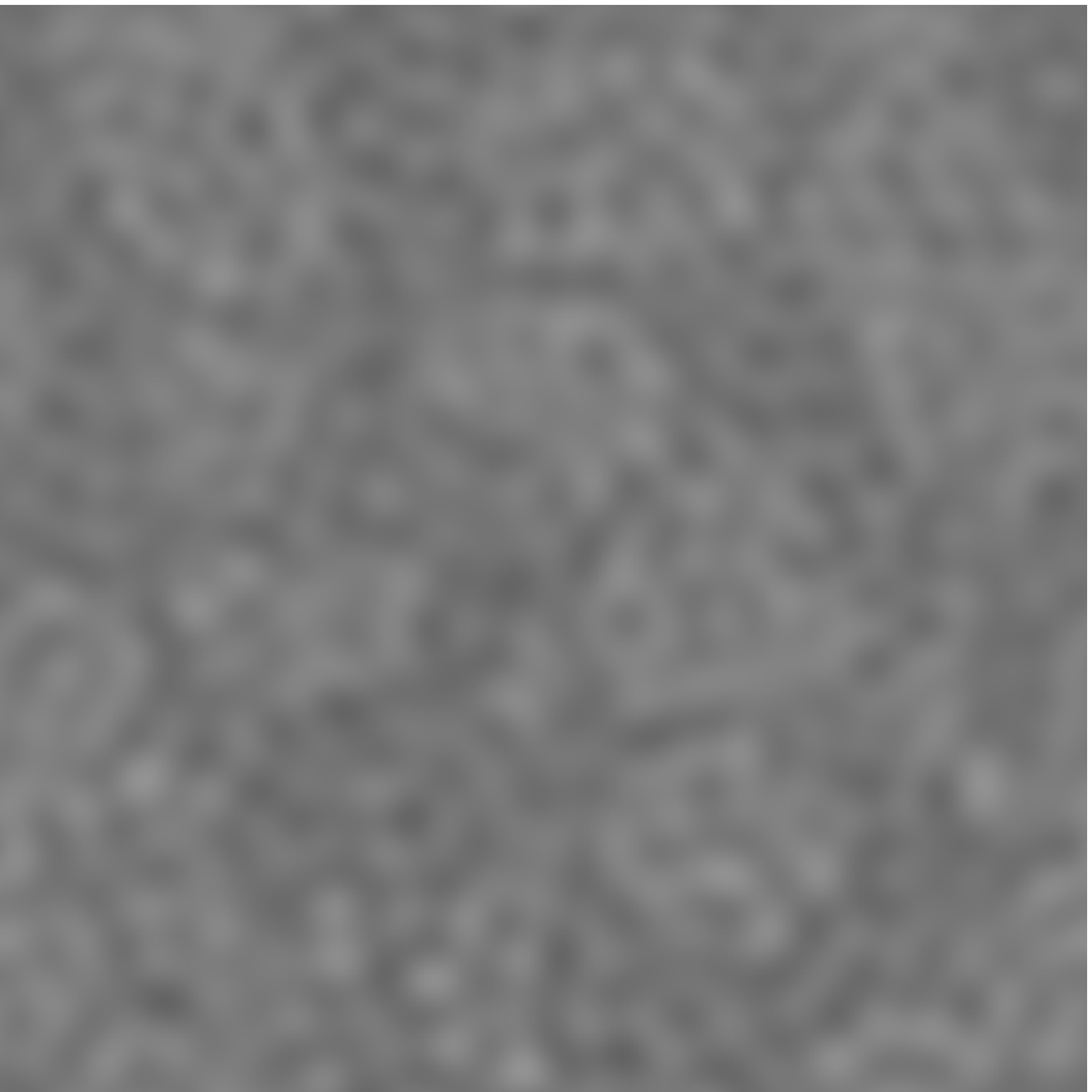}\\
(c)\includegraphics[width=6cm]{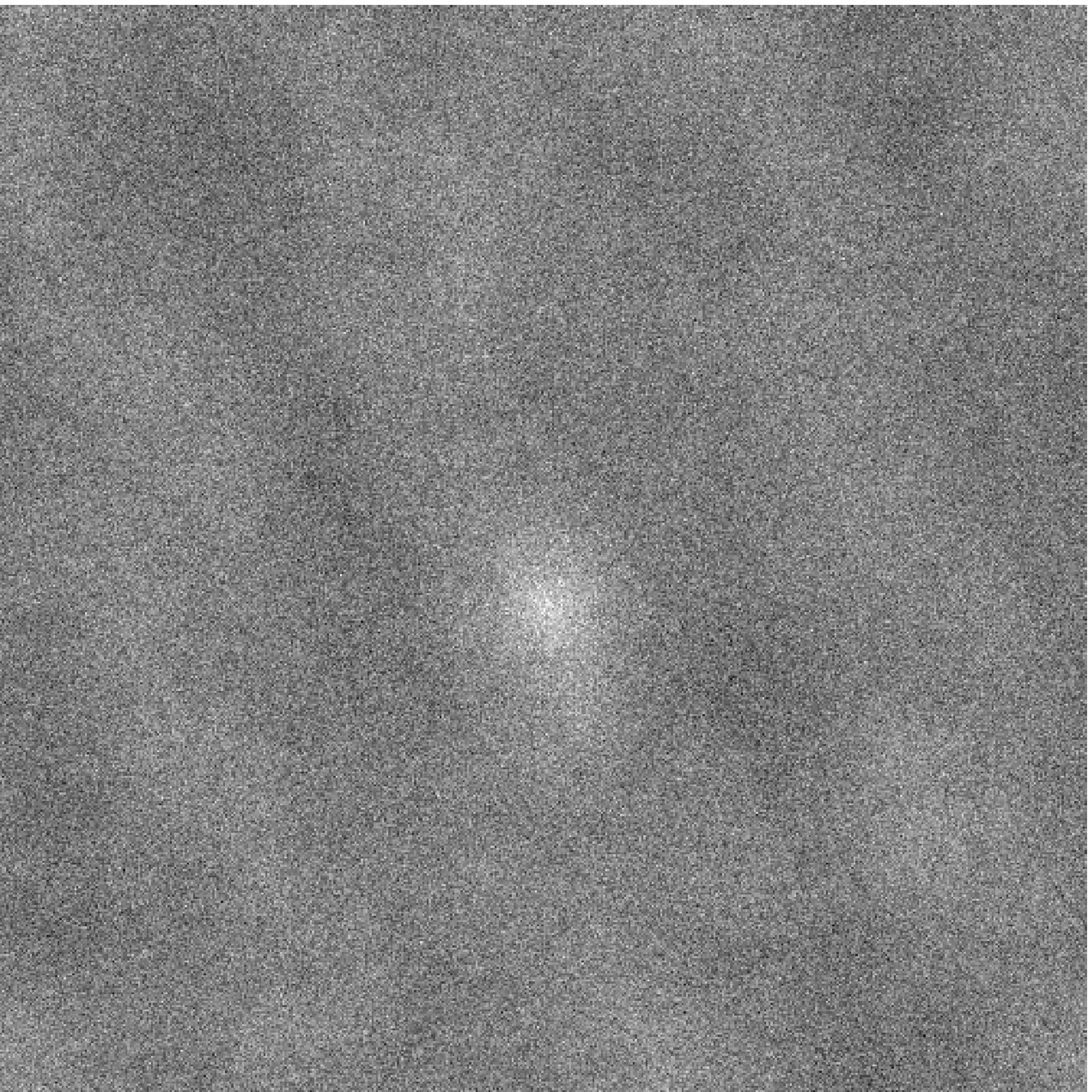}
(d)\includegraphics[width=6cm]{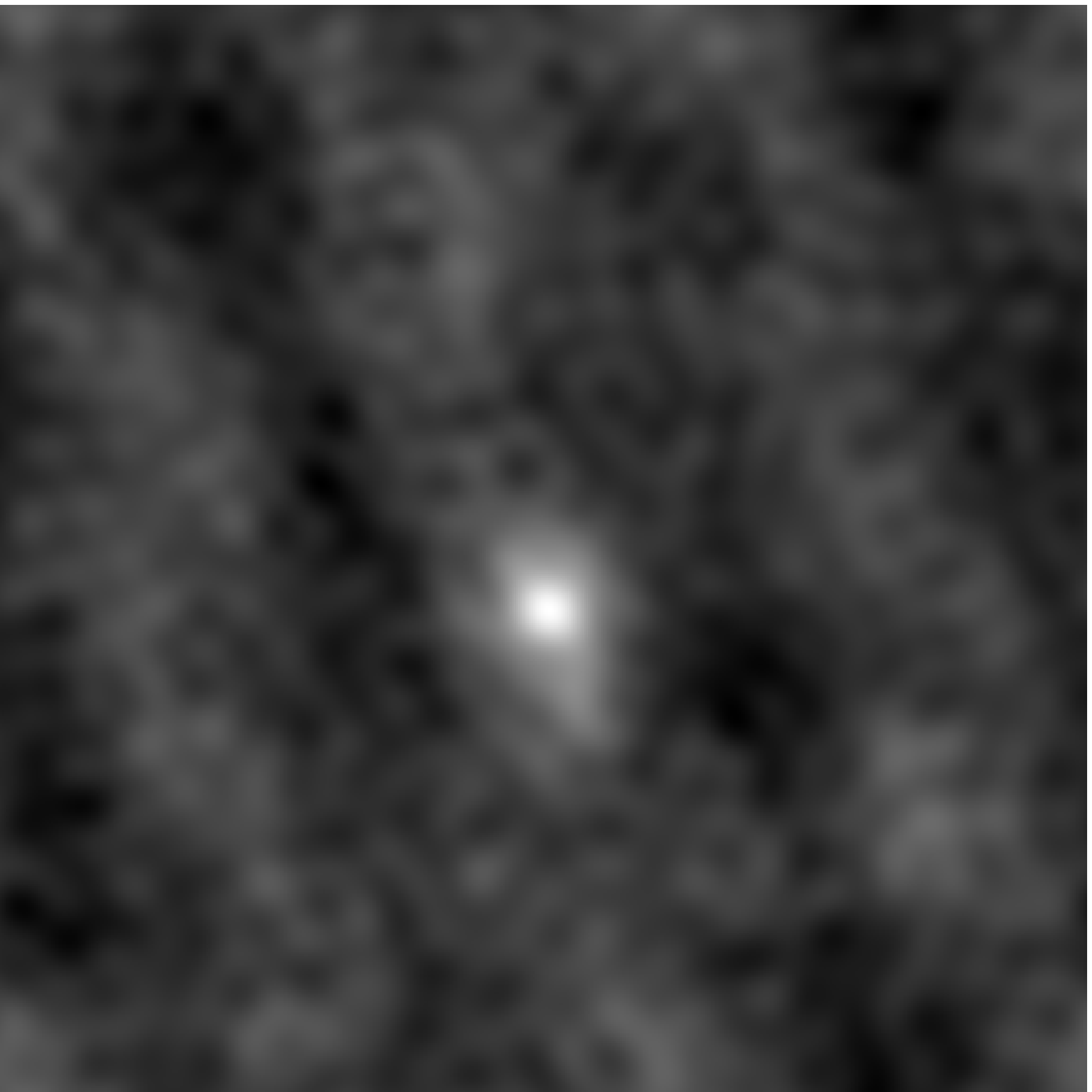}
\caption{Illustration of the cross-correlation displacement detection with noise filtering. 
(a)~Original frame, 
(b)~noise-reduced frame, 
(c)~unprocessed cross-correlation function, and
(d)~correlation function of two noise-reduced frames.
}
\label{fast_composition_fig_correlation_func}
\end{figure}

Several correction methods are being developed that compensate for these
effects. Some work on correcting the time-dependent drift
distortions has been performed in fields similar to scanning electron
microscopy \cite{kawasaki-drift}, \cite{mantooth}, \cite{chang-wang}, \cite{xu-li}. 
A research in drift-distortion evaluation and correction in SEMs has been
published in \cite{sutton1}, \cite{sutton2}, \cite{sutton3}. Technique described
in these papers covers correction in images with slow drift and low
magnification. The overall imaging times are high, reaching tens of minutes. The
magnification does not exceed 10000. 
Technique for very fast SEM or Scanning helium-ion beam microscopy,
where signal-to-noise ratio (SNR) may drop below $5\times10^{-1}$,
is still needed. This manuscript describes a possible correction method based on
composition of drift-distortion corrected SEM images. The technique uses cross-correlation for
displacement detection. It not only provides more accurate images, but also sample
position information, which can be successfully employed in diagnostic
applications. The method is implemented as a software program in the C
programming language. With this approach, the solution is fast, multi-platform,
multi-processor capable, and moreover can be easily integrated into the majority
of the SEM software.

\section*{Drift Effect on Images}
In the SEM, the image is formed by scanning over the sample
and acquisition of an intensity value at each location on the sample corresponding
to a pixel in the image. The intensity value $\xi(\vec{r})$ depends on the landing
position of the electron beam (on the sample) $\vec{r}$. 
Most SEMs use the raster pattern for scanning over the
sample. Let the raster pattern be defined by the time-dependent vector function:
\begin{eqnarray}
\vec r_r(t) &=& M\left(x(t)\vec e_x+y(t)\vec e_y\right),\\
t_p &=& t_D + t_d,\nonumber\\
y(t) &=& \left\lfloor\frac{t}{Xt_p+t_j}\right\rfloor,\\
x(t) &=&\left\lfloor\frac{t}{t_p}\right\rfloor - Xy(t),\\
0 \le &t& \le Y(Xt_p+t_j),\nonumber
\label{fast_composition_eq_raster}
\end{eqnarray}
where $t$ is time, $M$ is a constant of the length on sample corresponding to a
single-pixel step. $x$ and $y$ are column and row indexes in the SEM image.
$\vec{e_x}$ and $\vec{e_y}$ are the unit vectors in x- and y-direction, $t_D$ is
the dwell time of one pixel, $t_d$ is the dead time between two pixels, $t_j$ is
the time needed to move the beam to the beginning of the new line. $\lfloor q
\rfloor$ is a symbol for the ${\rm floor}(q)$ function as used in programming
languages. $X$
and $Y$ are the pixel-width and pixel-height of the SEM image. These equation are in agreement with those published
in \cite{sutton1}.

\begin{figure*}
(a)\includegraphics[width=6cm]{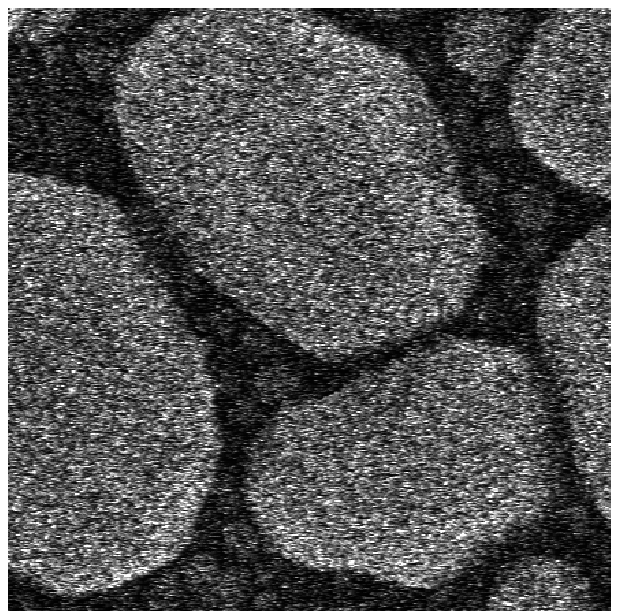}
(b)\includegraphics[width=6cm]{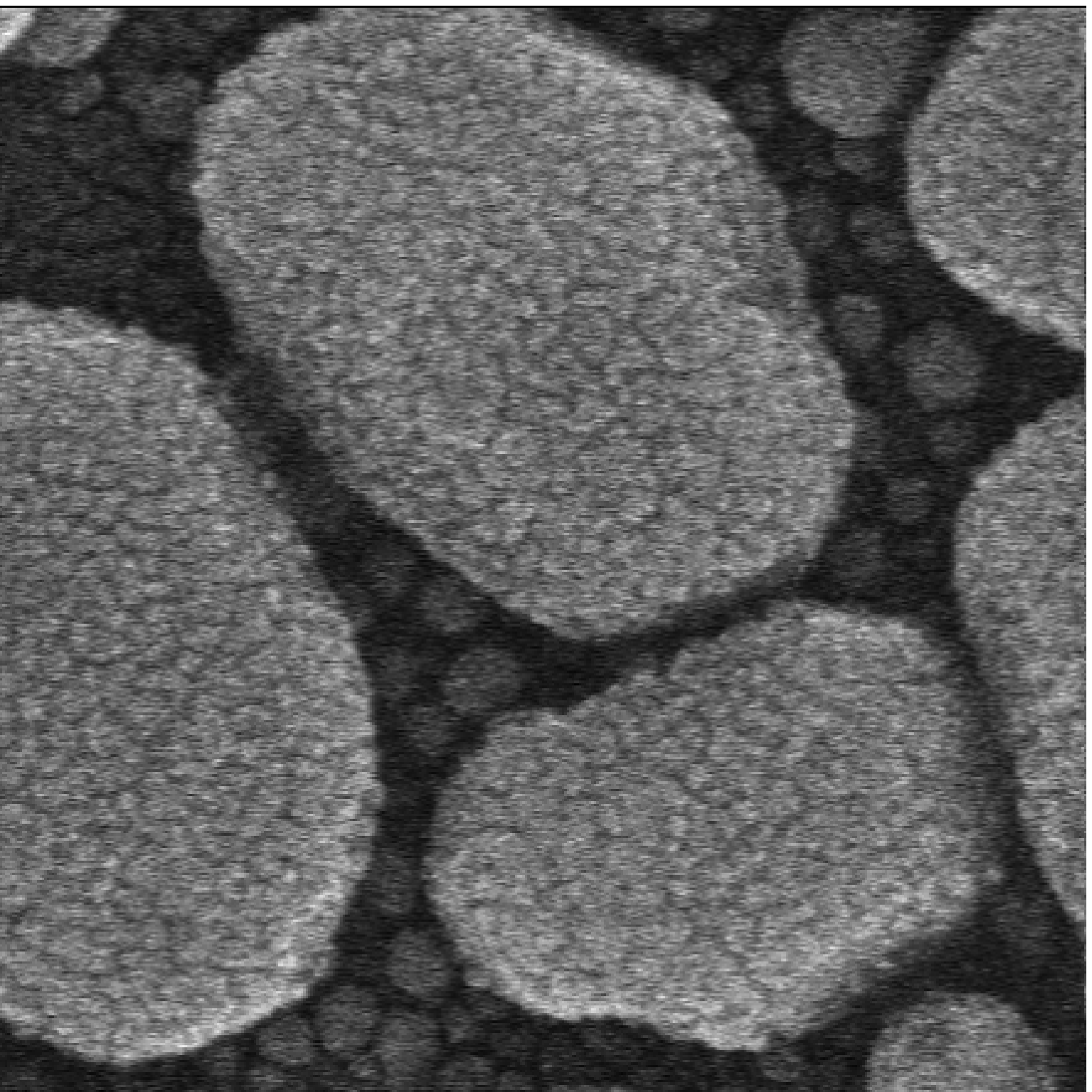}\\
(c)\includegraphics[width=6cm]{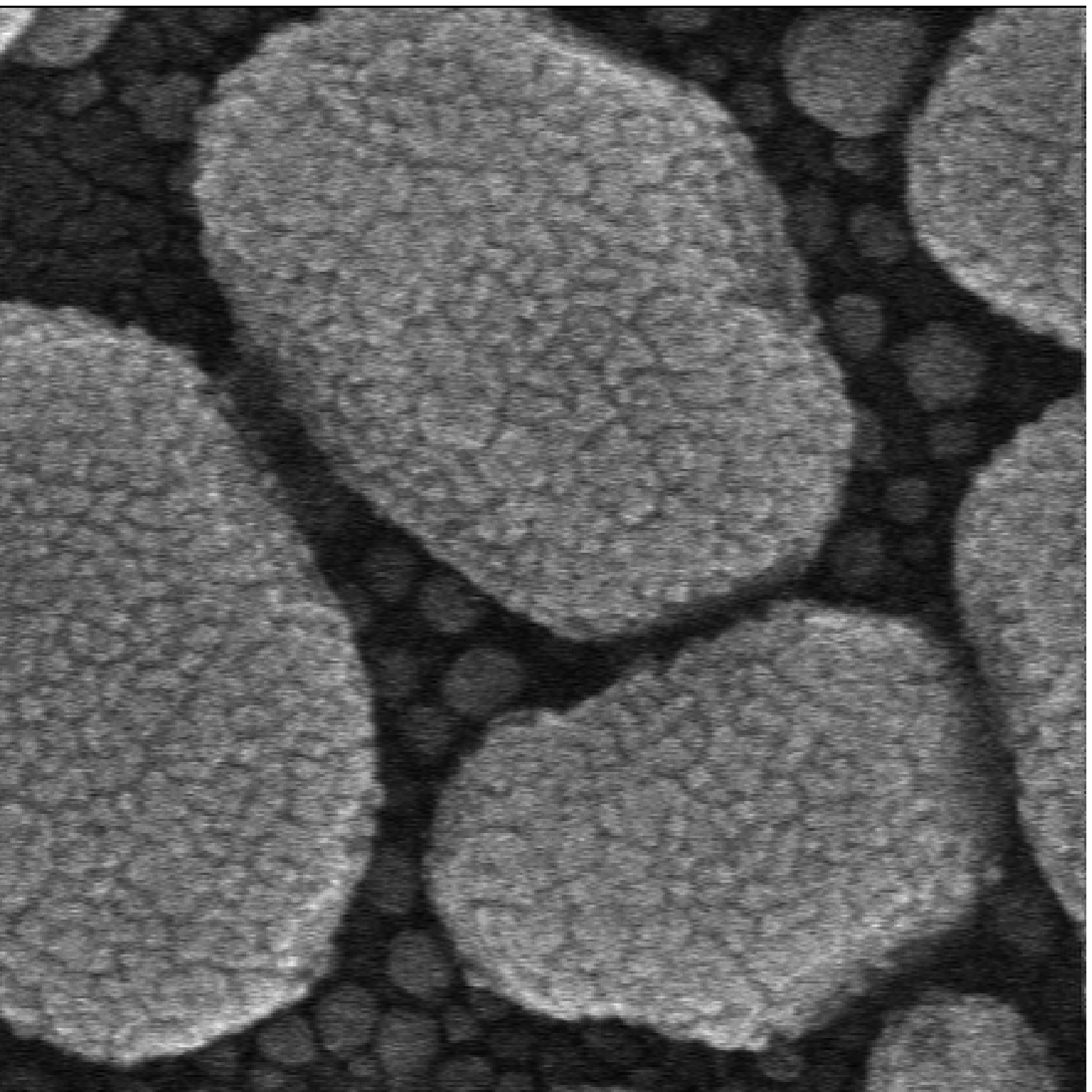}
(d)\includegraphics[width=6cm]{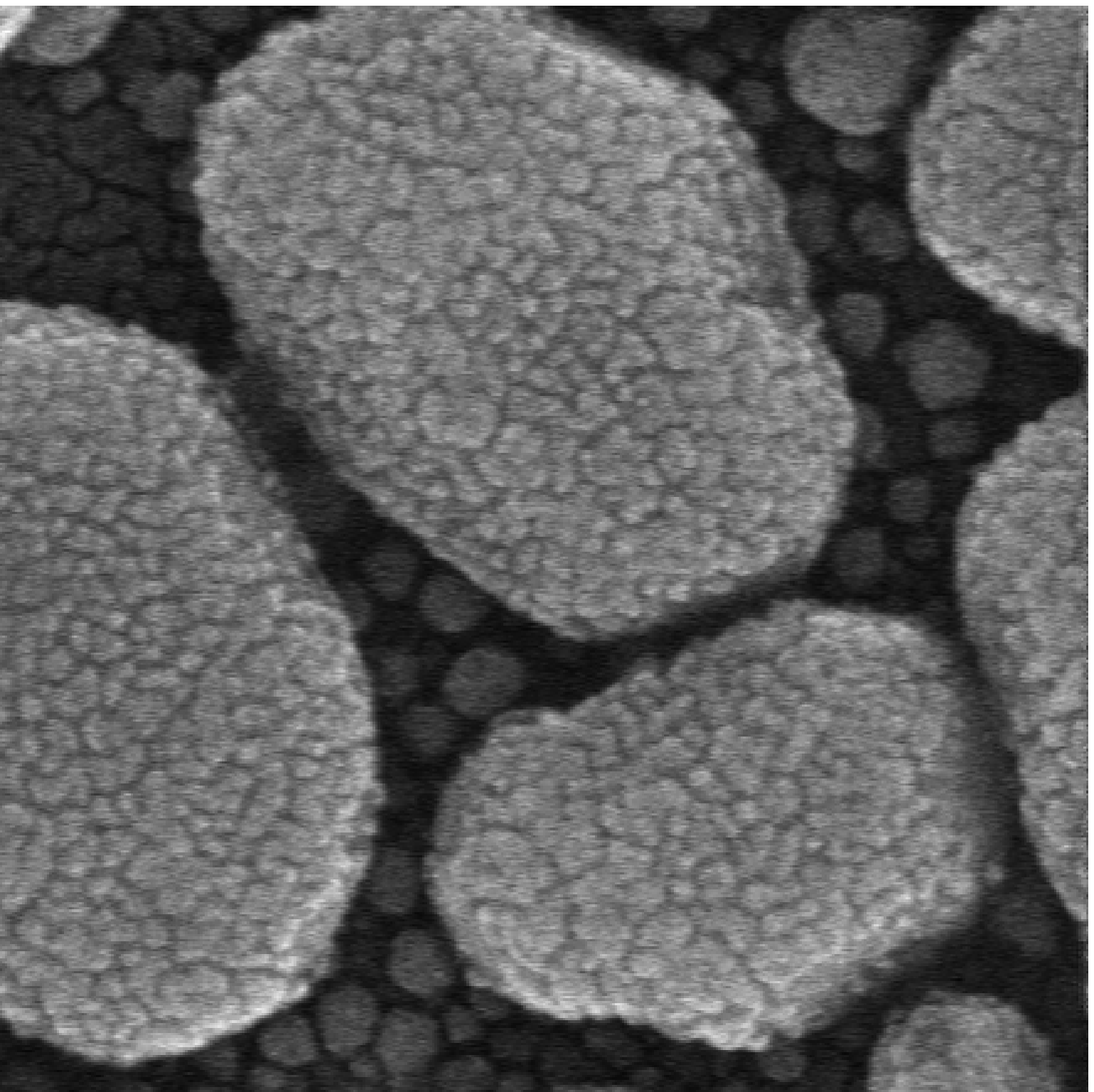}\\
(e)\includegraphics[width=6cm]{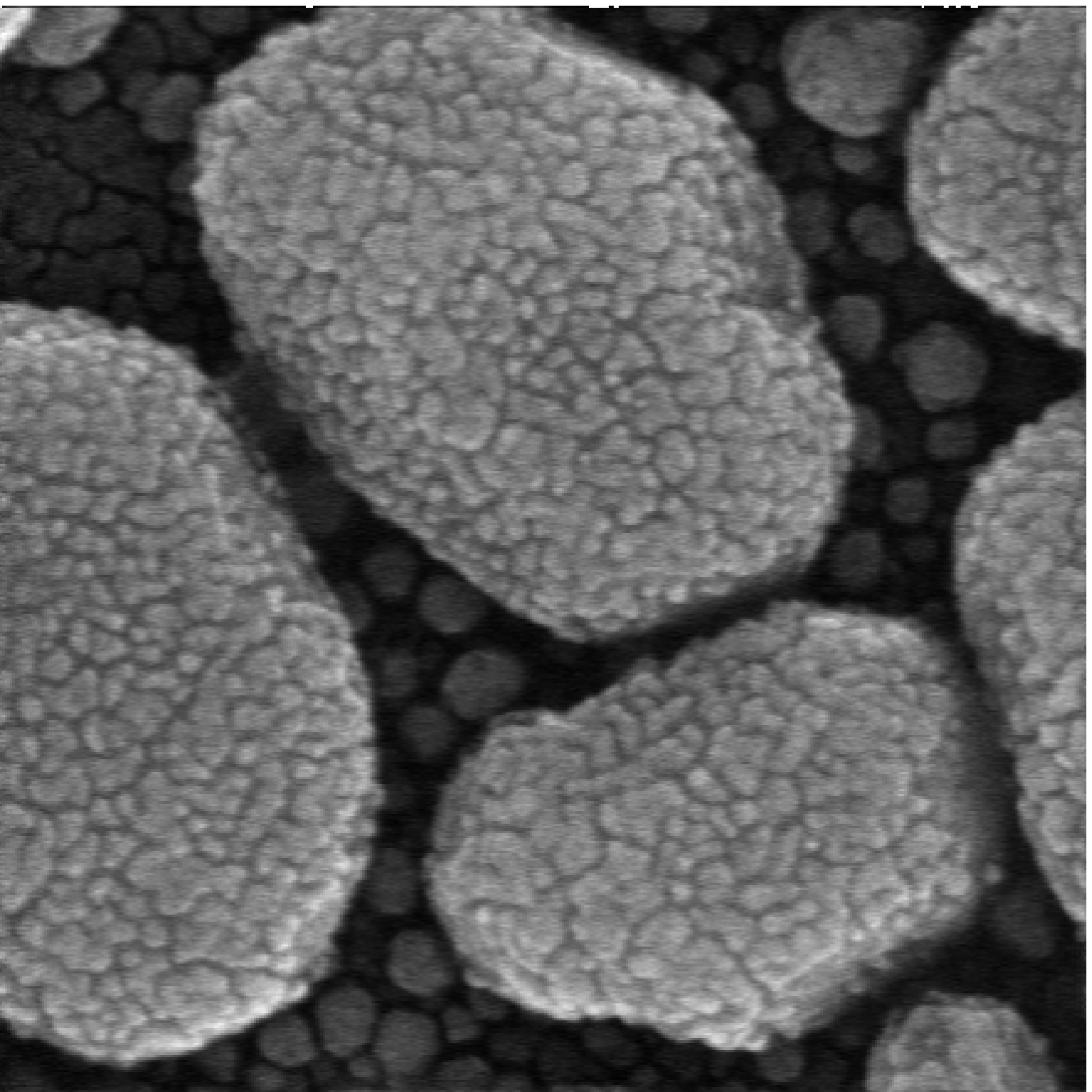}
(f)\includegraphics[width=6cm]{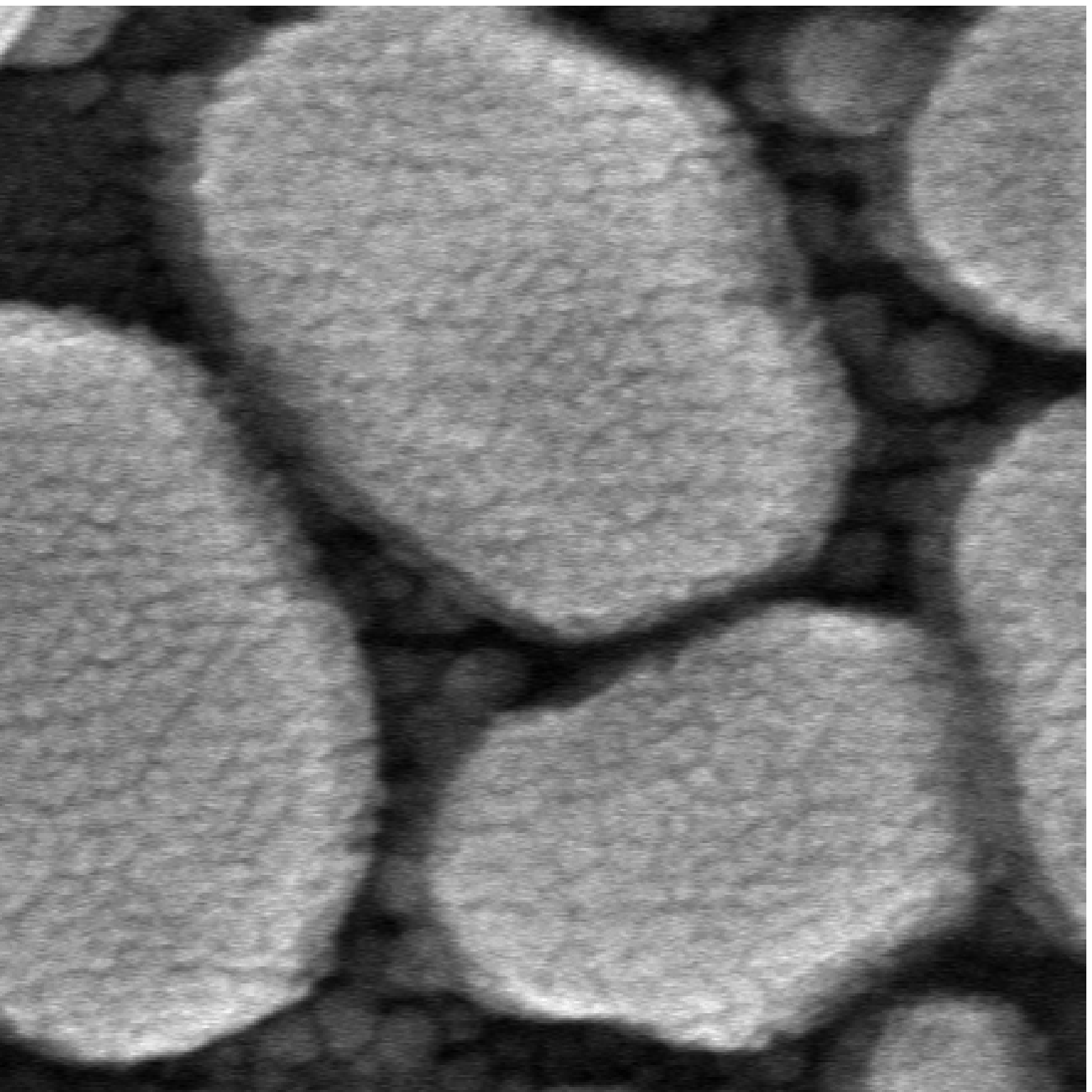}
\caption{Demonstration of the method on real SEM images of the gold-on-carbon
resolution sample. Horizontal field-of-view is 441~nm for all images, 
(a)~single acquired image with the pixel dwell time 50~ns, 
(b)~composition of 10 images, 
(c)~composition of 20 images, \
(d)~composition of 40 images, 
(e)~corrected composition of 120 images, and 
(f)~Plain average of the same 120 images.}
\label{fast_composition_fig_goldoncarbon_results}
\end{figure*}

Imaging in the SEM may be defined as a relation between the intensity map of the sample
$\xi(\vec{r})$ and the SEM image $I(x,y)$:
\begin{equation}
I(x(t),y(t)) = K\xi(\vec{r}(t)).
\label{fast_composition_eq_imaging}
\end{equation}
The relation between $I$ and $\xi$ may in practice be very general. For simplicity, let $K$
be a constant in this paper, since this does not affect generality of the
described technique.
In the ideal case $\vec{r}(t) = \vec r_r(t)$; however, drift and space
distortions are always present in scanning microscopes and
they often significantly affect the position $\vec{r}$:
\begin{equation}
\vec{r}(t) = \vec{r}_r(t) + \vec{D}_d(t) + \vec{D}_s(\vec{r}_r).
\label{fast_composition_eq_distortions}
\end{equation}
The space distortion $\vec D_s$ is constant in time and may be simply
compensated for, when its function is known. This kind of distortion is caused by
non-linearities in deflection amplifiers and appears mostly at
low magnifications. On the other hand, the drift distortion $\vec D_d$ is
changing in time, its function is usually unknown, and it most significantly
affects the high-magnification images. The drift distortion may arise from several
sources; e.g. translational motion of the sample, tilt or deformation of the
electron-optical column, outer forces and vibrations, or temperature expansion.
High-magnification images are very sensitive to drift distortion, since 
microscopic displacements, tilts, or temperature changes can easily cause
nanometer distortions and displacements, which can significantly impair the SEM image
and its usability for nanometer-scale measurements.

The drift-distortion function is generally unknown, however, since it
characterizes motion of physical bodies, it must be continuous and thus
square-integrable. Therefore, drift-distortion function may be 
expanded to Fourier series:
\begin{eqnarray}
D_{cd}(t) &=& \sum\limits_{n=-\infty}^\infty c_n{\rm e}^{-{\rm i}nt},\\
\vec{D}_d &=& \Re(D_{cd}) \vec{e}_x + \Im(D_{cd}) \vec{e}_y,\\ 
U &\propto& \sum_{n=-\infty}^\infty c_n^2 n^2,
\label{fast_composition_eq_fourexp}
\end{eqnarray}
where $c_n$ are the (complex) Fourier coefficients, $U$ is the overall energy of the
drifting system. 
Since $U$ is limited, for high $n$ the coefficients $c_n$ must be nearing zero.
In practice, $c_n$ for frequencies higher than 200~Hz correspond to noise only
and are negligible. This approximate number is based on experimental values. Therefore, the $D_{cd}(t)$ can be written:
\begin{equation}
D_{cd}(t) \approx \sum\limits_{n=-N}^N c_n{\rm e}^{-{\rm i}nt},\\
\label{fast_composition_eq_fourexplim}
\end{equation}
where $N$ represents the highest significant angular frequency.

\begin{figure*}
\includegraphics[width=13cm]{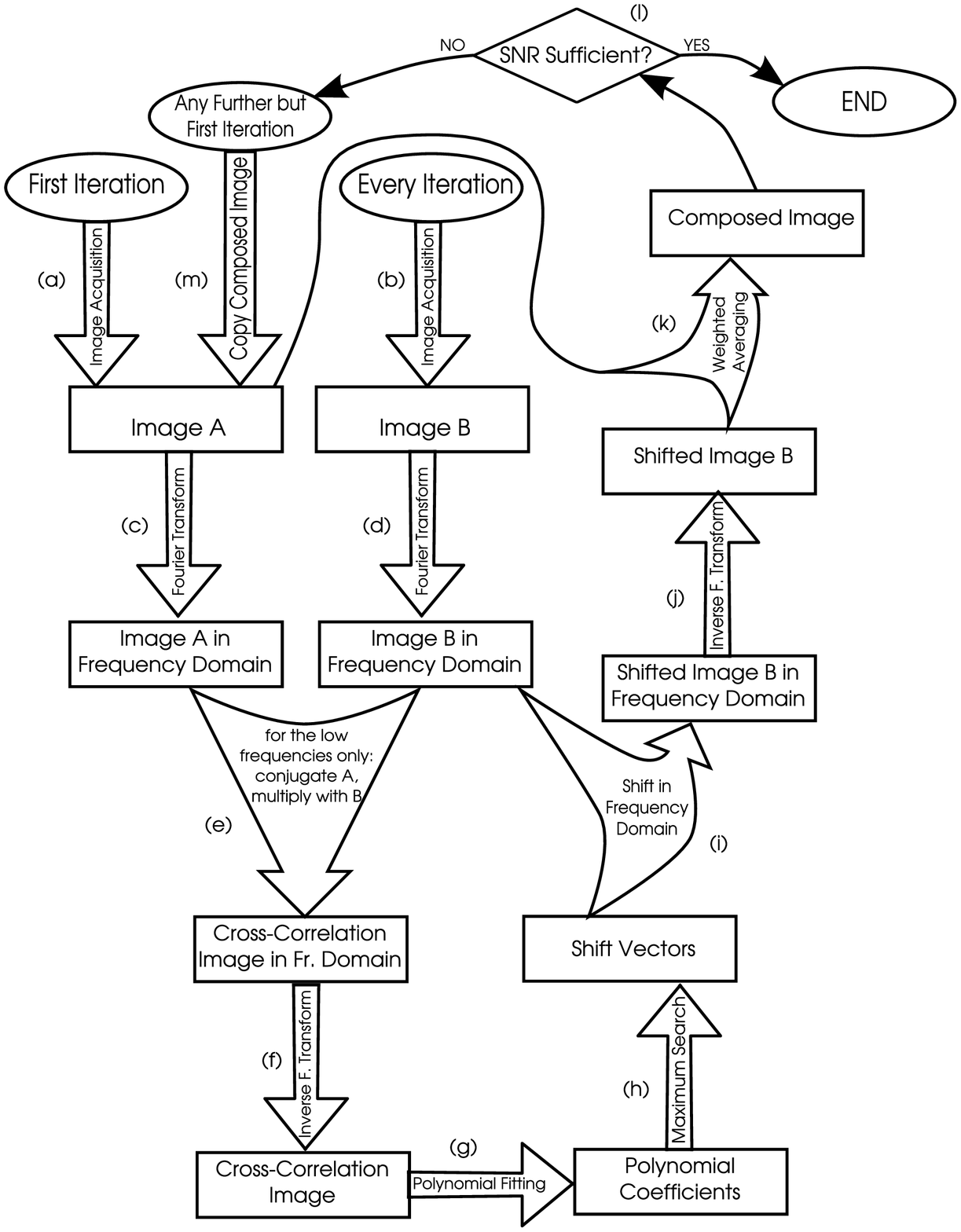}
\caption{
Illustration of the image composition procedure. Boxes denote entities like images
or numbers, arrows indicate processes. The letters (a)---(m) represent
individual steps in the procedure.
}
\label{fast_composition_fig_flowchart}
\end{figure*}

\section*{SEM Imaging Methods}
In the SEM, the acquired intensity signal always contains noise. The intensity
function is thus a superposition of real signal and noise:
\begin{equation}
\xi(\vec{r},t) = \xi_s(\vec{r}) + \xi_n(t),
\label{fast_composition_eq_inten}
\end{equation}
where $\xi_s$ is the position-dependent real signal and $\xi_n$ is the
time-dependent noise. This noise is a superposition of Poisson noise, originating
from the electron source and the secondary emission,  noise originating from
the amplifier and electronics, quantization-error noise, etc. Due to the central limit
theorem, it is legitimate to suppose that the mean value of this noise is zero:
\begin{equation}
<\xi_n(t)> = 0.
\label{fast_composition_eq_clt}
\end{equation}

In order to obtain a SEM image with a desired level of noise, the overall pixel
dwell-time $t_{D}$ must be sufficiently high. 
Unfortunately, the electron yield is usually low and the $t_{D}$ must often be
set to times ranging from tens to several hundreds of $\mu$s.

There are two
common techniques to achieve this in the SEM, i.e ``slow-scan'' and ``fast scan''.
With the ``slow-scan'' method,  the image is acquired within a single scan. The acquired value is in this case:
\begin{eqnarray}
I(x(t_0),y(t_0)) &=&
\frac{K}{t_D}\int\limits_{t_0}^{t_0+t_D}\left(\xi_s(\vec{r}(t))+\xi_n(t)\right)\d t,
\label{fast_composition_eq_intxi}\\
&&\int\limits_{t_0}^{t_0+t_D}\xi_n(t)\d t \approx 0,
\label{fast_composition_eq_noisereq}\\
\vec{r}(t)&=&\vec{r}_r(t_0) + \vec{D}_s(\vec{r}_r(t_0)) + \vec{D}_d(t)
=\nonumber\\
&=& {\rm const} + \vec{D}_d(t).
\label{fast_composition_eq_possingpix}
\end{eqnarray}
The noise is reduced by long integration as shown in
\eq~(\ref{fast_composition_eq_noisereq}). Required level of noise determines
the dwell-time $t_D$.  Since the desired beam position does not change during
the acquisition of a single pixel, the only changing component of the position
(\eq~(\ref{fast_composition_eq_distortions}))
is the $D_d$ as stated in \eq~(\ref{fast_composition_eq_possingpix}). In practice, the
drift-distortion-related displacements are not significant between two pixels,
because the time $t_p$ is still not long enough. However the line-acquisition
time $Xt_p+t_j$ may already be much larger than the period of the highest
frequencies.
Therefore, if the
``slow-scan'' technique is employed, the line-scans and thus also the images may
be significantly distorted (See \fig~\ref{fast_composition_fig_examples}). The distortion is time-dependent, and thus different
for each line and image and cannot be corrected, unless additional
information about the drift-distortion function $\vec{D}_d(t)$ is provided.

The other common imaging method in SEM is the ``fast-scan''. The image is
composed of multiple ($N_i$) frames, for which averaging is the mostly applied
technique. The
frames are acquired with the lowest possible pixel-dwell time $t_D$. Since, in
practice, the $t_D$ can be set to as low as 25~ns, the change in the
drift-distortion function during this time is negligible and the integral
(\ref{fast_composition_eq_intxi}) can be approximated as constant. The image pixel
value is then an average of corresponding frame-pixel values:
\begin{eqnarray}
I_k(x(t_0),y(t_0)) &=& K \xi_s(\vec r(t_0+kt_f))+\nonumber\\
 &+& K\xi_n(t_0+kt_f), 
\label{fast_composition_eq_fastscani}\\
I(x,y) &=& \frac{1}{N_i}\sum_{k=0}^{N_i} I_k(x,y).\\
t_f &=& Y(Xt_p+t_j)+t_{jj},
\end{eqnarray}
$t_f$ is a time period between beginnings of acquisition of two following
frames, $t_{jj}$ is the dead time between the end of acquisition of one frame and
beginning of the next one. Considering \eq~(\ref{fast_composition_eq_clt}), the
higher $N_i$, the lower noise level is present in the composed image. The required noise-level
thus determines the number of composed frames $N_i$. For high $N_i$:
\begin{equation}
\sum_{k=0}^{N_i-1} \xi_n(t_0+kt_f) \approx 0.\\
\end{equation}
Because the scanning raster pattern is constant for all frames,
\begin{equation}
\vec r_r(t_0+kt_f) = \vec r(t_0).\\
\end{equation}
\eq~(\ref{fast_composition_eq_fastscani}) may be expanded:
\begin{eqnarray}
I(x(t_0),y(t_0)) &=& \frac{K}{N_i} \sum_{k=0}^{N_i-1} \xi_s[\vec r_r(t_0) + 
\nonumber\\
&+& \vec D_s(\vec r_r(t_0)) + \vec D_d(t_0+kt_f)].
\label{fast_composition_eq_fastscani_exp}
\end{eqnarray}

With current SEMs, the frame-acquisition time $t_f$ can be much lower than
the period of even the highest drift-distortion frequencies. The
drift-distortion within the single-frame acquisition time is then minimal. 
However, it becomes significant during acquisition of the whole image,
especially, when the dead times $t_{jj}$ are high, which is the case
even with many current instruments. Considering drift effects negligible within a
single frame, the drift affects all image
pixels equally. Point-spread function (PSF) may be constructed from the
function $\vec D_d(t)$. Such a PSF consists of multiple separate points, which
produces blurry images similar to \fig~\ref{fast_composition_fig_shift_demo}.  The PSF can not be used for
deconvolution-based drift-distortion correction, since it is unknown like the
$\vec D_d$ itself.

\section*{Inter-Frame Drift-Distortion Correction}
The ``fast-scan'' method may be significantly improved using drift-distortion
correction. This is possible, when the frames are taken during short enough
times and $Y(Xt_p+t_j) \ll 2\pi/N$. Since the space-distortion $\vec D_s$ is much less
pronounced and much smaller that the drift-distortion $\vec D_d$ at very high
magnifications, it will be neglected from now on. The
\eq~(\ref{fast_composition_eq_fastscani_exp}) then becomes:
\begin{eqnarray}
I(x,y) &=& \frac{K}{N_i} \sum_{k=0}^{N_i-1} \xi_s[\vec r_r + \vec
D_{dk}],\\
\vec D_{dk} &=& \vec D_d(t_0+kt_f).
\end{eqnarray}
The image is in this case the mean value of $N_i$ displaced images. 

Fortunately, under certain conditions, it is possible to find the displacement
vectors of the images, which are equal to the drift-distortion values $\vec D_{dk}$.
These vectors then can be compensated for and thus the drift-distortion can be
corrected. One possible approach is a cross-correlation-based displacement
detection, which is used in this work. However, choice of the method determines the requirements, which may include low-enough noise,
well-pronounced image features, etc. The complete set of requirements will be
addressed in future publications.

\section*{Cross-Correlation with Noise Reduction}

If two image frames $f$ and $g$
contain similar features at different positions, the cross-correlation integral
has a large value at the vector corresponding to the displacement of the
features.  The SEM digital image frames are in this application represented by
discrete two-dimensional real functions. Therefore, the two-dimensional discrete
cross-correlation is applied. 

If the image frames are noisy, the peak in the cross-correlation function
becomes overridden by numerous other peaks, corresponding to random correlation
of noise (\fig~\ref{fast_composition_fig_correlation_func}c) This often makes
finding the displacement vector impossible. This could be tackled by low-pass
frequency filtering. This can be performed in the frequency domain. The cut-off
frequency is determined by the filter-radius $R$.  Although, this filter
significantly wipes out all high-frequency features from the image, and
therefore it is inapplicable for general reduction of noise, it still works very
well for the total-maximum search of a two-dimensional function.  The
maximum of the cross-correlation function becomes higher above the background
and it is easy to find it. (See
\fig~\ref{fast_composition_fig_correlation_func}d).

According to the cross-correlation theorem \cite{papoulis-cc}, the
cross-correlation can be calculated using the Fourier transform. The widely used
FFTW3 \cite{frigo-fftw3} algorithm is applied for Fourier transform calculations.
In order to speed up the calculation, the cross-correlation is combined with the frequency
filtering. (See \fig~\ref{fast_composition_fig_flowchart}e.) The conjugation and multiplication is done only in the central
circle (the lowest frequencies) of the Fourier image, while the rest is 
zeroed:
\begin{eqnarray}
J &=& \left\{ 
\begin{array}{l l}
  [{F}(f(\vec r))]^*\cdot{F}(g(\vec r)) & \quad \mbox{if $|\vec r| \le R$,}\\
  0 & \quad \mbox{otherwise.}
\end{array} \right.\\
I &=& F^{-1}(J)
\label{}
\end{eqnarray}
Then, the inverse Fourier transform is applied and the noise-reduced
cross-correlation image is obtained. 
 For every pair of frames, only two forward and one inverse Fourier
transforms are needed, while one of them is also part of the next
pair. 

\begin{figure}[h!]
(a)\includegraphics[width=6cm]{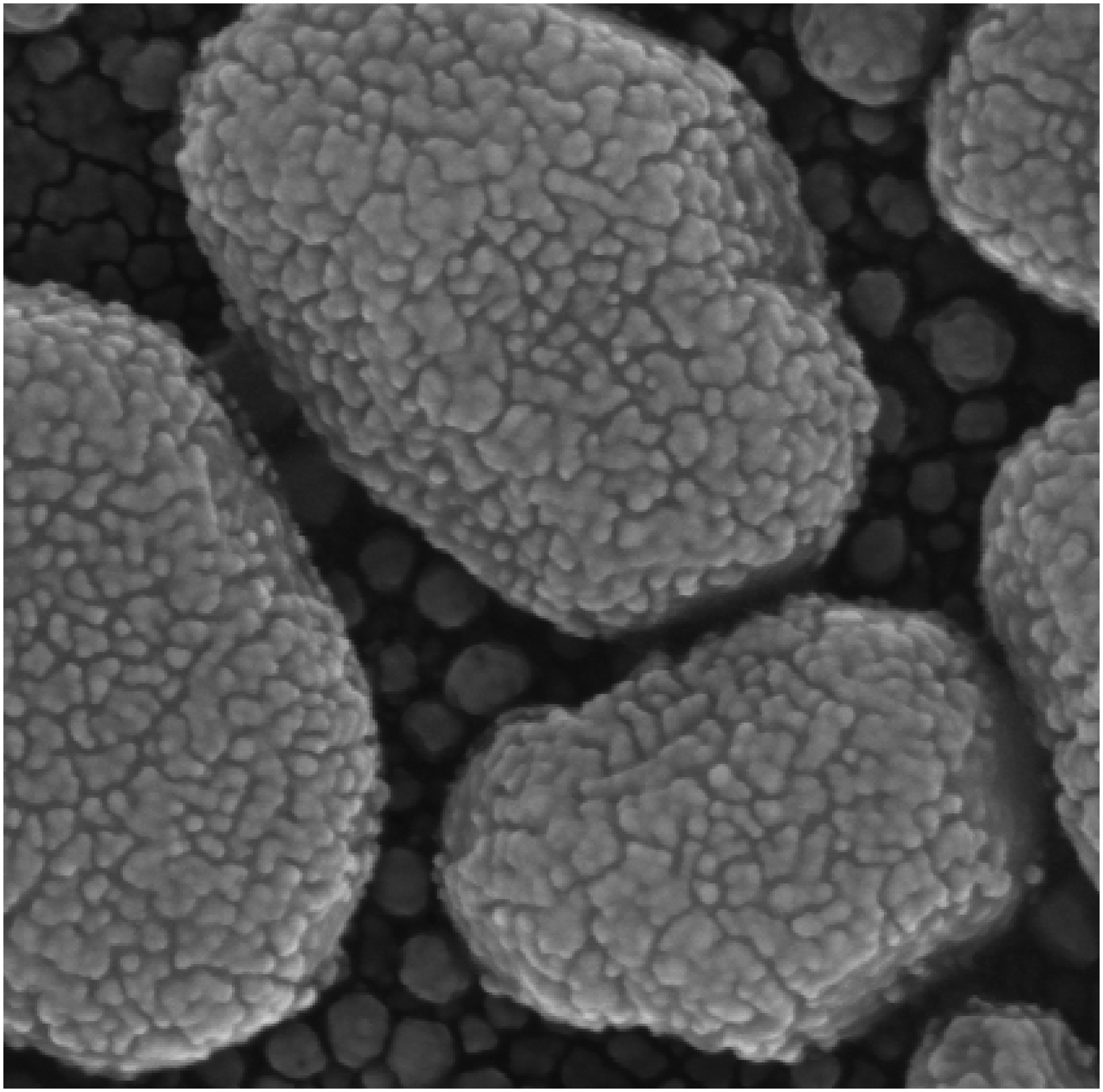}
(b)\includegraphics[width=6cm]{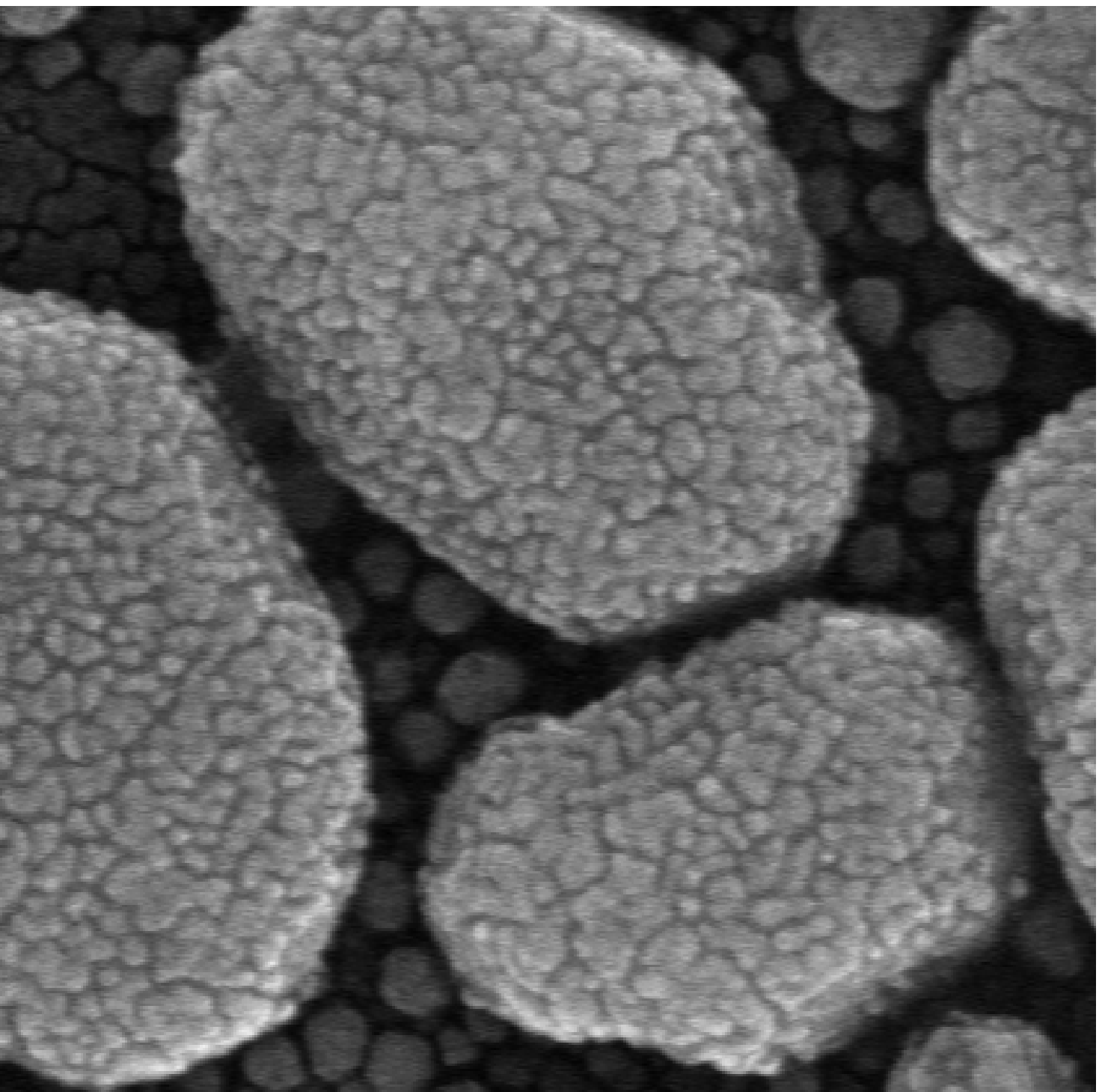}\\
(c)\includegraphics[width=6cm]{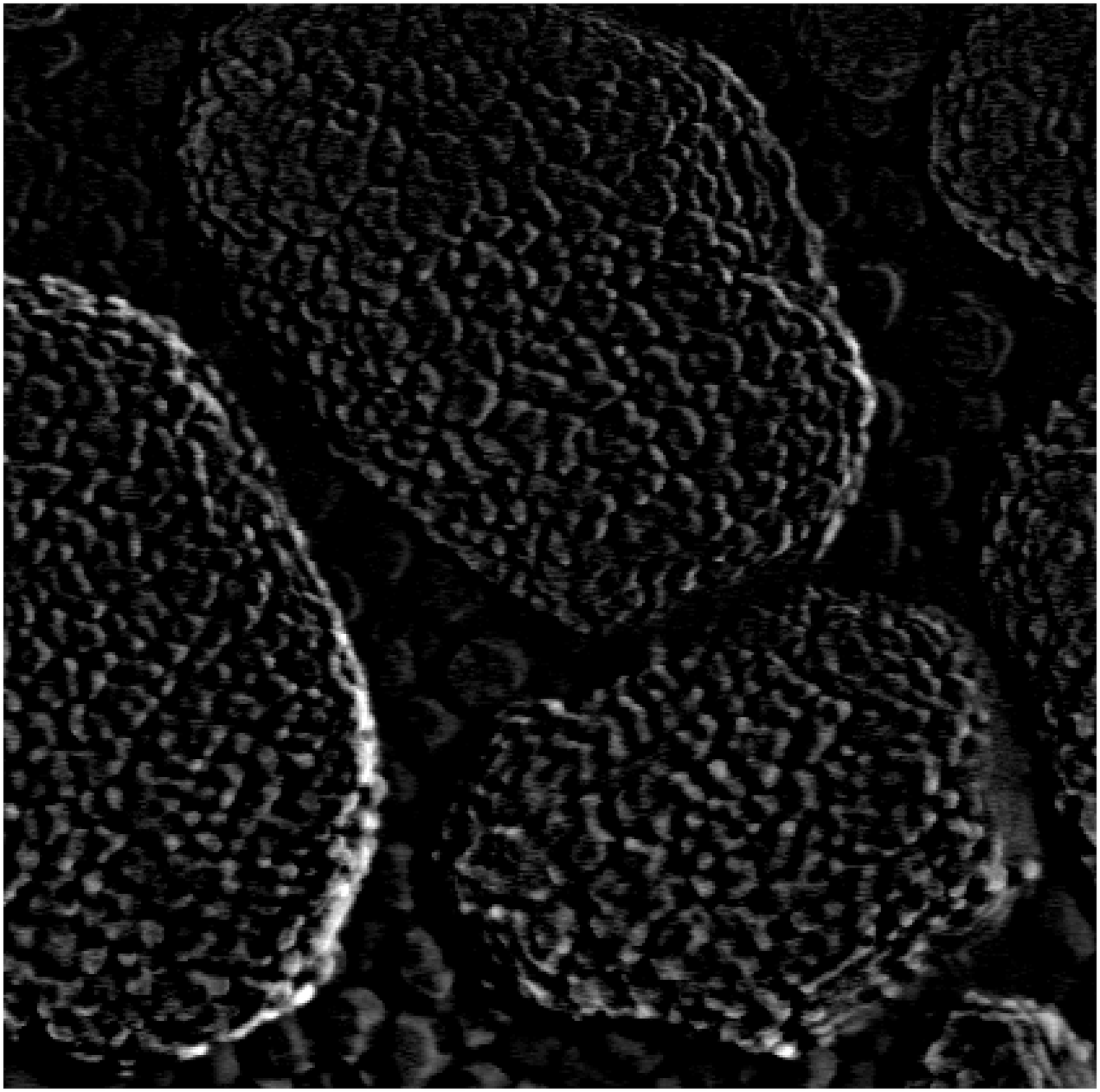}
\caption{
The results of the corrected composition compared with the uncorrected ``slow
scan''
demonstrated on a real SEM image of a gold-on-carbon resolution sample.
Horizontal field of view is 422~nm. 
(a)~Slow-scan image with the pixel dwell time 300~$\mu\rm{s}$. 
(b)~The result with the new technique. 
(c)~Difference between the two images.}
\label{fast_composition_fig_goldoncarbon_results_comparison}
\end{figure}

\section*{Sub-Pixel-Accuracy Displacement Detection}

In order to find the displacement vectors,  the maximum of the cross-correlation is
searched for. Since this method is sample-dependent, there may appear problems
finding the displacement vector corresponding to the transition. For example, if the
image contains periodic features, there are several maxima in the
cross-correlation image. In case of large blur or noise, the peak may be very
wide and the uncertainty in the position of the maximum may be very high.

Plain search for maximum provides just single-pixel accuracy of the displacement-vector.
The peak may be interpolated by a suitable function, which enables for
calculating the displacement vectors with sub-pixel accuracy. The third-order
two-dimensional polynomial function with coefficients $k_0$---$k_9$ was chosen
as a suitable function for the interpolation.
The coefficients may be found by polynomial fitting using the
common least squares method, which is widely used in similar applications. 
The Cholesky factorization \cite{gentle-cholesky}
is applied to speed the calculation. The maximum is then numerically found; its position is the searched
displacement-vector with sub-pixel accuracy. However, the exact accuracy-value
calculation is not covered in this article.

Since the displacement-vectors are calculated with sub-pixel accuracy, it is
reasonable
to shift the images with sub-pixel accuracy as well. Let the
displacement-vector be $\vec{s} = s_x \vec e_x +  s_y \vec e_y$. The shift can
be performed with sub-pixel accuracy using the Fourier Transform, because
\begin{equation}
F_s(\omega, \phi) = F(\omega,\phi)\exp(-i\omega x_s-i\phi y_s).
\label{fast_composition_eq_2d_fourier_shifted}
\end{equation}

The obtained displacement-vectors with the sub-pixel accuracy obtained in the previous step are
used for shifting, which compensates for the inter-frame drift-distortion. The corrected
images are then pixel-by-pixel averaged together, however, other composition
methods, e.g. median filtering, can be also used.

\begin{figure*}
(a)\includegraphics{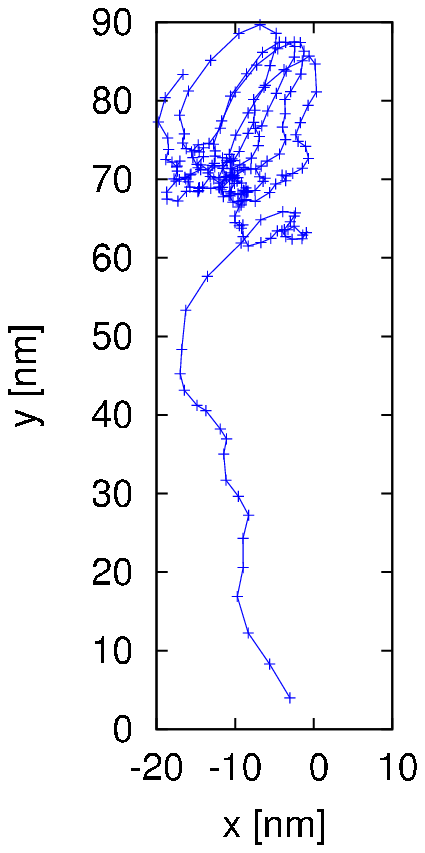}
(b)\includegraphics{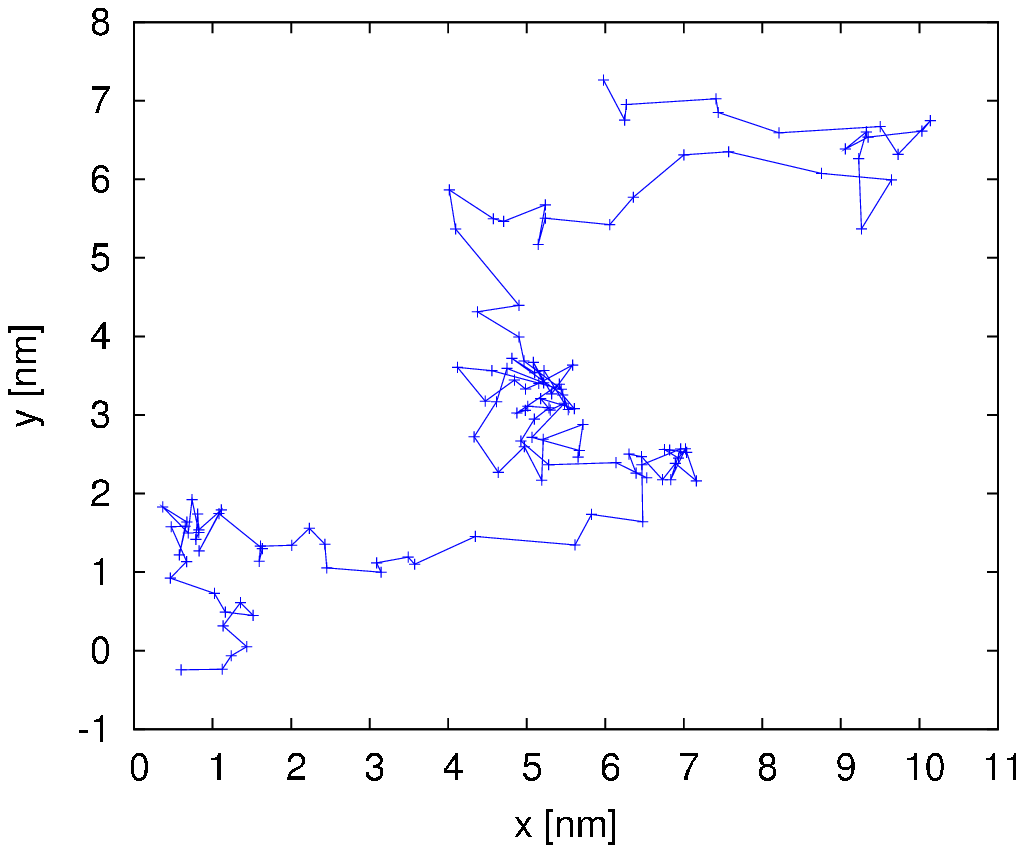}
\caption{ Example of a drift tracking. Subsequent displacement vectors are 
displayed in the graph. 
(a)~Drift-distortion of a sample placed on a fixed stage. Frames were taken every 60~s. 
(b)~Drift distortion of the sample used in
\fig~\ref{fast_composition_fig_goldoncarbon_results}. Frames were taken every
1~s.}
\label{fast_composition_fig_position_tracking}
\end{figure*}

\section*{Composition Procedure}
The procedure of the method is illustrated in
\fig~\ref{fast_composition_fig_flowchart}. The process starts with acquisition
of two frames; A and B (\figs~\ref{fast_composition_fig_flowchart}a and b). In
order to minimize the influence drift-distortion  on the frames,
these must be taken with minimum pixel-dwell time possible, which is usually limited
by the instrument. Both frames are then converted into the frequency domain (\figs~\ref{fast_composition_fig_flowchart}c and d).
These frequency-domain images are then conjugated and multiplied, which is combined with frequency
filtering
(\fig~\ref{fast_composition_fig_flowchart}e). This results in a
cross-correlation image in the frequency domain. 
Then, the cross-correlation image in space domain is obtained
(\fig~\ref{fast_composition_fig_flowchart}f). The cross-correlation image is
interpolated by the third-order two-dimensional polynomial function. This enables finding the displacement
vectors with sub-pixel accuracy (\fig~\ref{fast_composition_fig_flowchart}g).
The coordinates of the maximum denote the found displacement vector
(\fig~\ref{fast_composition_fig_flowchart}h). This displacement vector is used to shift
the image B in its frequency-domain representation, which enables the sub-pixel
accuracy alignment (\fig~\ref{fast_composition_fig_flowchart}i). Shifted image B
is converted into the space domain
(\fig~\ref{fast_composition_fig_flowchart}j) and averaged with the image A
(\fig~\ref{fast_composition_fig_flowchart}k). Image A has (except in the first
iteration) higher information weight, as it already represents a sum of multiple
image frames. If the SNR
is not sufficient (\fig~\ref{fast_composition_fig_flowchart}l), the composed
image is copied into the frame A (\fig~\ref{fast_composition_fig_flowchart}m)
and a new frame B is acquired. The process then repeats until the SNR
is sufficient, or the software runs out of frames.

\section*{Results}

The discussed image-composition method was tested on gold-on-carbon resolution
images (\fig~\ref{fast_composition_fig_goldoncarbon_results}) and on artificial
images. A Mac Pro
computer with two dual-core Intel Xeon Central Processor Units (CPUs) and 4~GB of
Random Access Memory (RAM) was employed for the calculations. The 64-bit edition of Gentoo Linux
Operating System (OS) was installed on the computer.

For the real images, the pixel dwell time was
set to the lowest instrument setting (50~ns). The frame-rate was 1 frame per
second, which was also the fastest setting. Single acquired frame
(\fig~\ref{fast_composition_fig_goldoncarbon_results}a) was very noisy; only the
most prominent features (about 200~nm in diameter) were clearly visible. A
composition of 10 frames
(\fig~\ref{fast_composition_fig_goldoncarbon_results}b) already contained visible
features in the background (about 20~nm in diameter); some inner structure of
the grains (sized about 5-10~nm) became
observable. Compositions of 20
(\fig~\ref{fast_composition_fig_goldoncarbon_results}c) and 40
(\fig~\ref{fast_composition_fig_goldoncarbon_results}d) frames embodied some more
detail. The inner structure of the grains, as well as all the background
features are clearly visible. The
composition of 120 frames from the new composition method
(\fig~\ref{fast_composition_fig_goldoncarbon_results}e) and the existing
averaging method (\fig~\ref{fast_composition_fig_goldoncarbon_results}f) are
included for a comparison of the new and the traditional averaging methods. The
traditionally averaged image was significantly more blurred than the image 
composed using the described method. Both images
have similar SNR. The final image
(\fig~\ref{fast_composition_fig_goldoncarbon_results}e) exhibits low noise and
high detail whilst preserving the shapes and dimensions.

The slow-scan image is displayed in
\fig~\ref{fast_composition_fig_goldoncarbon_results_comparison}a. The pixel
dwell time was 300~$\mu$s, which is a common choice for such imaging. The image
looked clean and noise free on inspection; however, the difference
between the
slow-scan image and the image obtained with the new composition method, as shown
in \fig~\ref{fast_composition_fig_goldoncarbon_results_comparison}c exhibited a
significant difference, which is believed to be associated with the distortions
in the slow-scan image. The image was acquired for 6000 times longer time than
the image in \fig~\ref{fast_composition_fig_goldoncarbon_results}a and the
drift-distortion affected the shapes significantly.

The obtained sequences of displacement-vectors were also used to track the sample
position with respect to the beam
(\fig~\ref{fast_composition_fig_position_tracking}). This information was very
usable for drift investigation. In the case of the sample of the fixed stage,
displayed in \fig~\ref{fast_composition_fig_position_tracking}a, there was a roughly 70-nm-long straight start-up
drift followed by a periodical circular drift, which was caused by periodical temperature
changes inside the electron-optical column.

On the other hand, the \fig~\ref{fast_composition_fig_position_tracking}b shows the
displacement-vector sequence associated with a typical drift in the SEM.
Assuming the obtained curve to be associated with a relative trajectory of
physical bodies with a position noise superimposed to it, it is possible to
estimate the accuracy of the displacement-vector searching to be approximately 0.5~nm,
which corresponds to 0.5~pixels.

Speed of calculation is another important aspect of this method and its implementation.
It was not possible to try the technique in a real-time imaging
application, since this would require integration of the technique into the SEM
software, which was not possible. However, the calculation times were measured
and on 512$\times$512-pixel-large images, a single search for a
displacement-vector took in average 0.08~s, while the times of individual frame
compositions are very consistent.

\section*{Conclusion} 
The technique is implemented as a
computer program written in C language, which is the advantage due to its optimization possibilities and
ease of possible incorporation into SEM software. On reasonably fast computers,
this program is capable of real-time processing. The algorithm is well
distributable, thus, it is suitable for running on computer clusters or
multi-core or multi-processor environments, including graphics processing units
(GPUs).  The method
has been verified on real and artificial SEM images demonstrating its usability for
true-shape imaging and for drift investigation applications. It was also tested
for the calculation speed, which is high enough for real-time processing, when
integrated into the SEM software.

Since the power of this method strongly depends on many factors, e.g.
sample-feature shapes, noise, image size, etc., its limits should be throughly
examined. Calculation of accuracy and confidence intervals, influence of
sample charging and contamination are still under investigation.
These issues will be addressed in future works on this project. 

\bibliographystyle{plain}
\bibliography{literature}


\end{document}